\documentclass[10pt,letterpaper]{article}
\usepackage{opex3}
\usepackage{color}
\usepackage{multirow}
\usepackage{subfig}
\begin{document}

%%%%%%%%%%%%%%%%%% title page information %%%%%%%%%%%%%%%%%%
\title{Impact of the slit geometry on the performance of wire-grid polarisers}

\author{Gwenaelle M\'{e}len,$^{1,*}$ Wenjamin Rosenfeld$^{1}$ and Harald Weinfurter$^{1,2}$}

\address{$^1$Ludwig-Maximilians-Universit\"at, 80799 M\"unchen, Germany\\
$^2$Max-Planck-Institut f\"ur Quantenoptik, 85748 Garching, Germany}

\email{$^*$gwenaelle.vest@lmu.de} %% email address is required

% \homepage{http:...} %% author's URL, if desired

%%%%%%%%%%%%%%%%%%% abstract and OCIS codes %%%%%%%%%%%%%%%%
%% [use \begin{abstract*}...\end{abstract*} if exempt from copyright]

\begin{abstract}
Wire-grid polarisers are versatile and scalable components which can be engineered to achieve small sizes and extremely high extinction ratios. Yet the measured performances are always significantly below the predicted values obtained from numerical simulations. Here we report on a detailed comparison between theoretical and experimental performances. We show that the discrepancy can be explained by the true shape of the plasmonic structures. Taking into account the fabrication details, a new optimisation model enables us to achieve excellent agreement with the observed response and to re-optimise the grating parameters to ensure experimental extinction ratios well above 1,000 at 850$\,$nm.
\end{abstract}
%150 wordsa

 % REPLACE WITH CORRECT OCIS CODES FOR YOUR ARTICLE, MINIMUM OF TWO; Avoid using the OCIS codes for “General” or “General science” whenever possible.

%%%%%%%%%%%%%%%%%%%%%%% References %%%%%%%%%%%%%%%%%%%%%%%%%

%%%%%%%%%%%%%%%%%%%%%%%%%%  body  %%%%%%%%%%%%%%%%%%%%%%%%%%

%optical counterparts of regular microwave polarizers.
\section{Introduction}
\indent Although the concept of microwave polariser based on subwavelength gratings dates back to the 19$\,^{th}$ century, it could not be adapted to the optical domain until the development of microfabrication technologies in the 1960s \cite{Bird1960}. Since then, these so-called Wire-Grid Polarisers (WGP) have been used frequently as they offer small footprints and large acceptance angles, and can now be found in daily-life devices such as sensors \cite{Shishido2011} or displays. Current research focuses onto low-cost and high volume manufacturing processes, UV applications \cite{Wang2014}, as well as high performance.  \\
 \indent A WGP consists of parallel metal stripes on a transparent substrate, such that the period $p$ is smaller than the wavelength of the incident radiation. In this configuration, the Transverse-Magnetic wave (\textit{TM}, also $\pi$ or $p$), polarised orthogonally to the stripes, can experience Extraordinary Optical Transmission (EOT) due to coupling to surface plasmon polaritons \cite{Lochbihler1994,Thio2001} and waveguiding effect through the slits \cite{Astilean2000,Takakura2001}. The subwavelength dimensions guarantee zero-$th$ order diffraction. On the contrary, the Transverse-Electric polarisation (\textit{TE}, also called $\sigma$ or $s$) is almost perfectly reflected. This leads to high polarising efficiencies, characterised by the Extinction Ratio, \textit{i.e.} the ratio of the transmission of orthogonal polarisations $ER = T_{TM}/T_{TE}$. \\
 \indent  Usually a trade-off has to be found between high ER and high transmittance of the \textit{TM} modes. Our application for quantum cryptography \cite{Vest2014} requires an array of four equivalent polarisers with a minimum ER of 1,000 at 850$\,$nm, while the overall transmission is not a critical parameter. Numerical simulations typically used to find this compromise promise very high ER which are, however, never measured in real devices. Discrepancies of several orders of magnitudes have been observed \cite{Wang2014,Ahn2005,Cetnar2012}. To achieve better experimental results, new structures such as dual-gratings have been proposed \cite{Yang2007}. 
  \\
\indent Here we show that geometrical deviations of the manufactured structure with respect to those used in the simulations are the reason for this discrepancy \cite{Siefke2014}. In particular our gold gratings exhibit trapezoidal stripes instead of rectangular ones, a common problem, with only a few exceptions \cite{Liu2006,Oh2007}, to all fabrication techniques such as anisotropic etching, Focused Ion Beam (FIB) milling and nanoimprint. We observe an approximately exponential decrease of the extinction ratio with the tilting angle of the slits. Yet accounting for the real shape of the wires, in particular for the average slit width, leads to very good agreement with the observed performances. Our improved numerical simulation model predicts that high extinction ratios can be achieved even with imperfect structures. By reducing the design slit width we obtain an experimental ER of better than 1,000 and a transmission of 9$\,$\% for a wavelength of 850$\,$nm. \\
\indent This paper is organised as follows: we first compare the grating parameters and the simulation results based on perfectly rectangular structures  (Section 2) with the performance of samples fabricated accordingly, exhibiting high discrepancies with the numerical computation (Section 3). In Section 4 we describe a refinement of our simulation model and an optimisation step to achieve the desired performances. Finally we confirm our new model with the characterisation of new optimised samples.

\section{Standard grating optimisation}

\begin{figure}[!h]
\centering
\includegraphics[width=.45\textwidth]{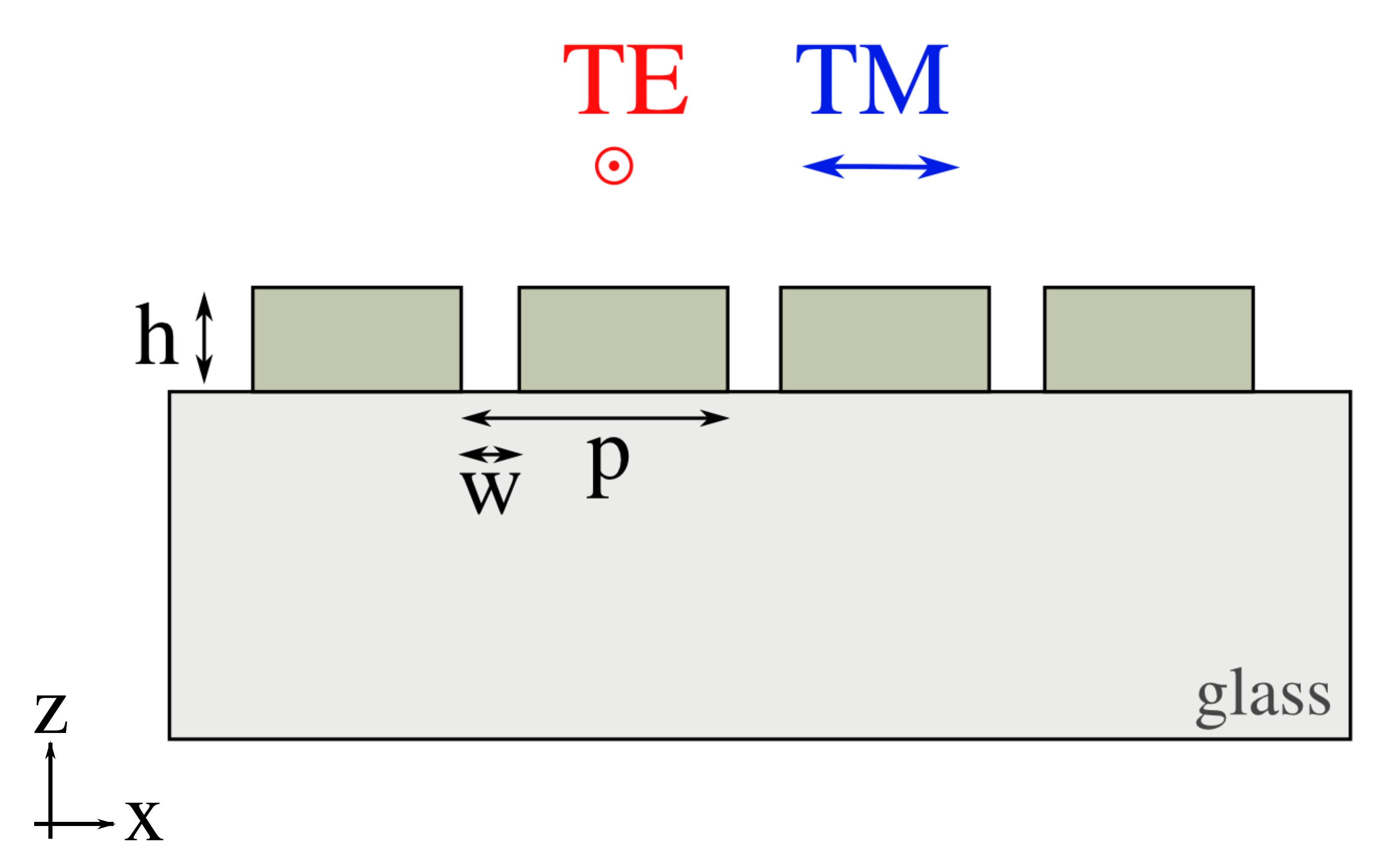} 
\vspace{-5pt} 
\caption{Geometry of a gold wire-grid polariser on a glass substrate.}
\label{geo}
\end{figure}

Considering a gold grating on a glass substrate, the parameters to be optimised are the period $p$, slit width $w$ and thickness $h$, as depicted in Fig. \ref{geo}. The period as well as the duty cycle (ratio $w$ to $p$) are directly related to the coupling of the \textit{TM} polarisation with surface plasmon on the top and on the bottom of the metal stripes. Certain combinations of parameters lead to efficient field confinement between the stripes (Rayleigh-Wood resonances), thereby increasing the resulting ER. In general smaller slits tend to a favourable increase of the polarisation selection for a fixed period, but also lead to lower \textit{TM} transmission. The latter is nonetheless also dependent on the metal height, which should be selected carefully to guarantee efficient tunnelling through the slit via mode-matched Fabry-P\'{e}rot cavity resonances. Increasing thickness leads however to an exponential decay of the transmission of the \textit{TE} polarisation, mainly responsible for the evolution of the ER. The interplay between both vertical and horizontal resonance phenomena is studied in \cite{Marquier2005,collin2002}. Yet due to fabrication limitations, high quality high aspect ratio structures are hard to manufacture, imposing a constraint on the grating parameter set.\\

\begin{figure}[!h]
\centering
\captionsetup[subfigure]{labelformat=empty}
\subfloat[\label{hw}]{\includegraphics[width=.47\textwidth]{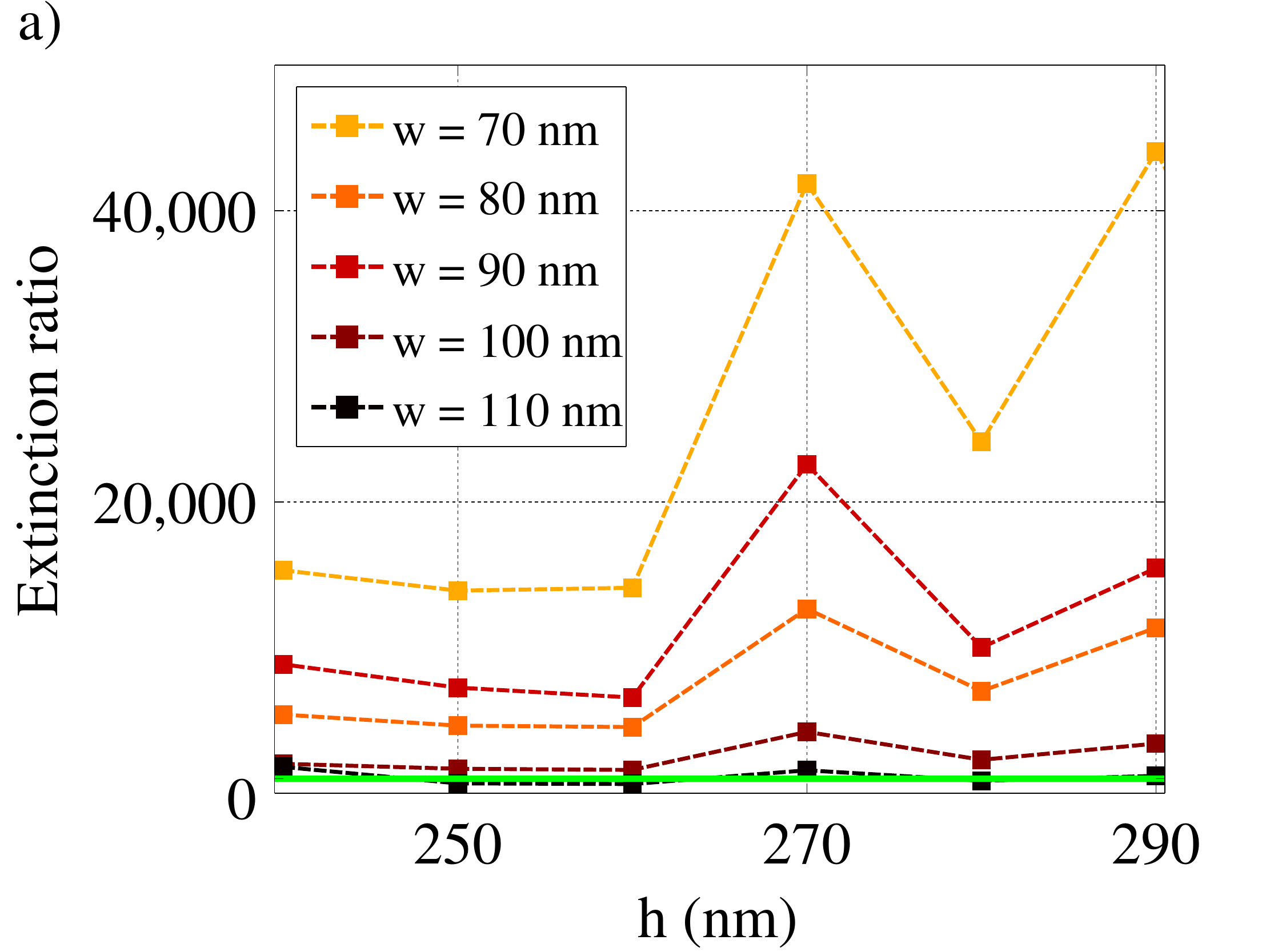}}
\hfill 
\subfloat[\label{terect}]{\includegraphics[width=.46\textwidth]{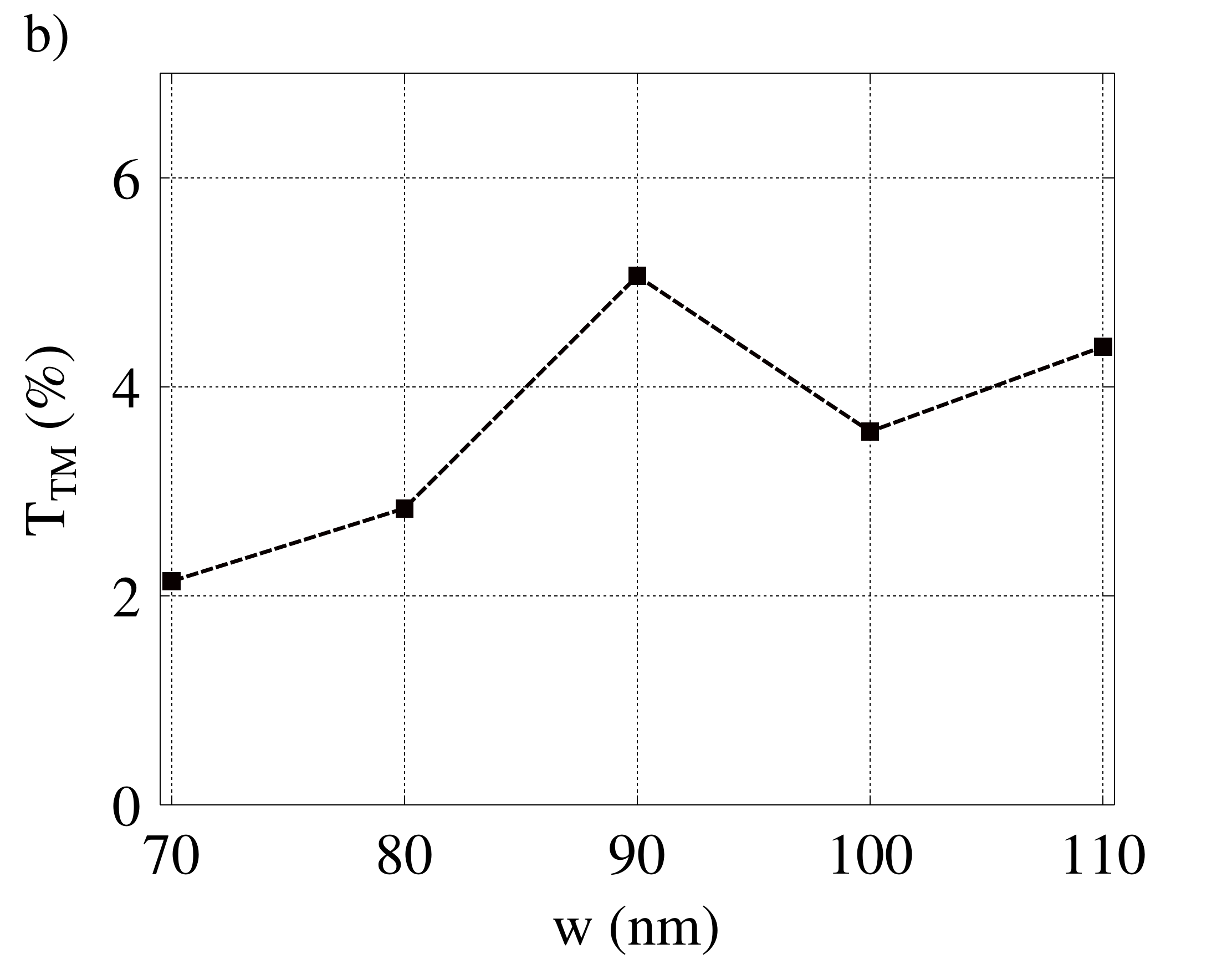}}
\vspace{-15pt}
\caption{Influence of the design parameters on the performance of the polariser. (a) Dependence of the extinction ratio on $h$ and $w$ for $p= 500\,$nm. Here the effect of the thickness is considered only around the first vertical resonance. The green line represents our design rule of $ER=1000$. (b) Transmission of the TM-modes as a function of $w$ for $h= 270\,$nm.  }

\end{figure}

Previous studies \cite{Guillaumee2009} indicate typical grating parameters, and a period of $p= 500\,$nm seems to deliver good performances around $ \lambda =850\,$nm. In this regime the condition $p \ll \lambda$ is not fulfilled, therefore the Effective Medium Theory (EMT) cannot be used. The optimisation of the slit width as well as the gold thickness was thus studied using FDTD simulations (MEEP \cite{ArdavanF.Oskooi2010}). This simulation program uses the Lorentz-Drude dispersion model with optical constants obtained from thin films measurements \cite{Rakic1998}. Due to the very long computation time associated with a resonant structure, the spatial resolution was limited to 7$\,$nm. Figure \subref*{hw} shows the dependence of the extinction ratio on the thickness and the slit width. Beyond $h=320 \,$nm, the ER reaches very high values but we rather concentrate on the first vertical resonance at $h = 270 \,$nm, as the fabrication of very narrow slits with straight flanks becomes more difficult with increasing metal thickness. We also observe on Fig. \subref*{terect} a horizontal resonance at $w=90 \,$nm, where the ER increases due to higher transmission of the \textit{TM}-modes.

\section{Fabrication and characterisation}

A 170$\,$\textmu m thick glass substrate was coated with 3$\,$nm Titanium (adhesion layer) followed by 265$\,$nm gold using Electron Beam Physical Vapour Deposition. Deposition rates were maintained relatively low (below 1.5$\,\mathring{A} \cdot s^{-1}$) to minimise the granularity and the surface roughness of the layers. Although simulations indicate a resonance for $h=270 \,$nm, better performances were experimentally observed for $h=265 \,$nm. This shift was also predicted by additional high-resolution simulations taking into account the presence of the adhesion layer. The theoretical ER obtained for the experimental configuration is nevertheless comparable with the one resulting from the simple model of gold ($h=270\,$nm) on a glass plate. The subwavelength gratings (120$\,  \times\,$120$\,$\textmu m) were then engraved using the Focused Ion Beam (FIB) milling technique (Zeiss Cross-Beam) at the Center for Nanostructures and Nanomaterials (ZNN, Munich, Germany). The wire-grid polarisers were characterised by measuring the ER for well-defined input polarisations. While this study focuses on performance at 850$\,$nm, we refer the reader to \cite{Siefke2014} for similar numerical and experimental results including wavelength dependence.\\

\begin{table}[!h]
\renewcommand{\arraystretch}{1.3}
 \centering
   \caption{Comparison between simulated and experimental extinction ratios for different slit widths. Here the structure is simulated with a period $p = 500 \,$nm a gold thickness of $h = 270 \,$nm, but fabricated with $h = 265 \,$nm with additional 3$\,$nm Ti.}
   \vspace{-5pt}
 \begin{tabular}{ c c c c}
Sample & $w$ (nm) & ER (experimental)  & ER (simulated) \\ \hline  \hline
$A_1$ & 120 & 380    &  \multirow{2}{*}{2520}  \\ %fib13 ter 

$A_2$ & 120 & 650   &      \\   \hline 
$B_1$ & 80  & 720   &  \multirow{2}{*}{12700}  \\ 
$B_2$ & 80  & 850   & \\
 \end{tabular}

   \label{oldsamples}
\end{table}

\begin{figure}[!h]
\centering
\captionsetup[subfigure]{labelformat=empty}
\subfloat[]{\label{st} \includegraphics[width=.27\textwidth]{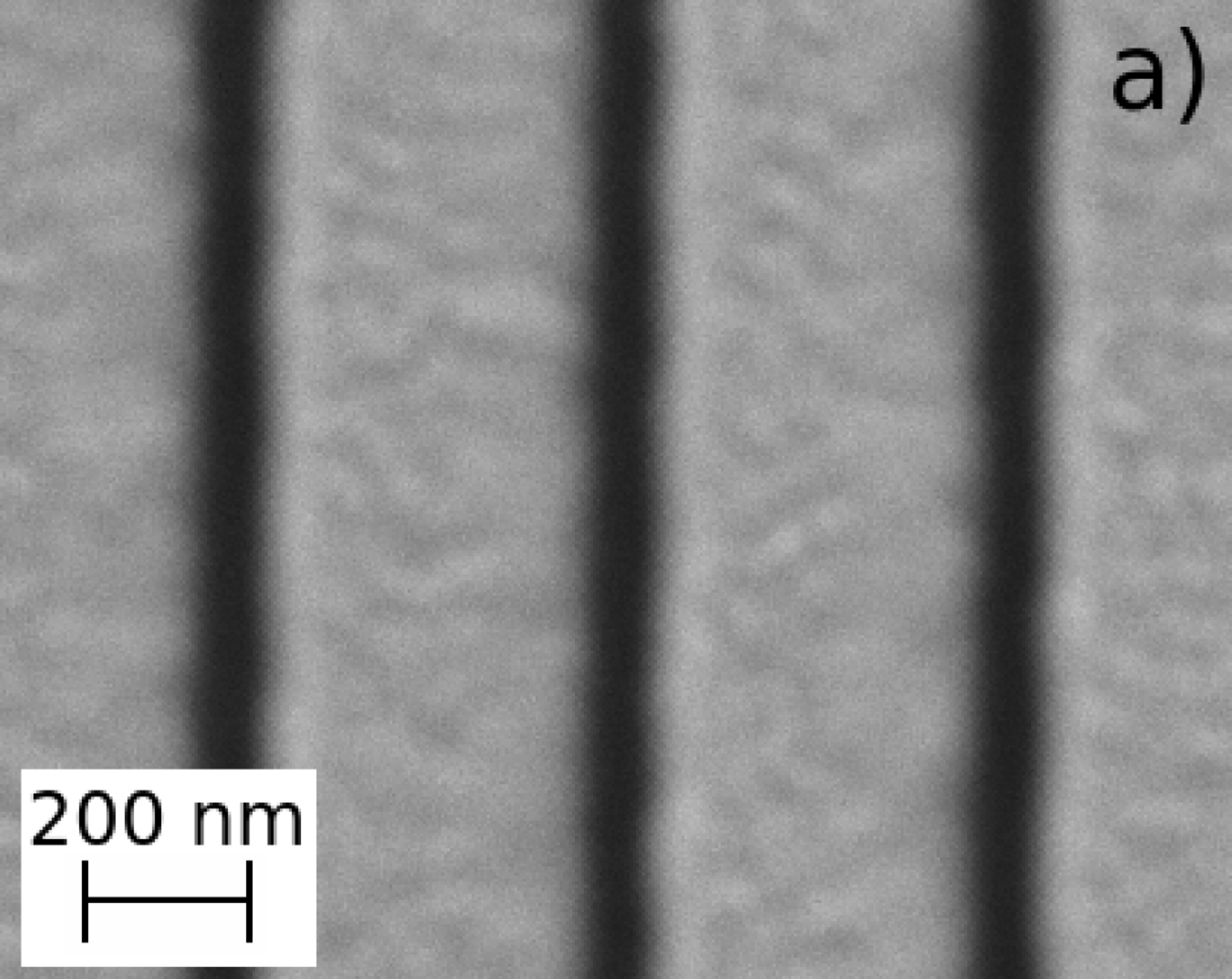}} \hfill
\subfloat[]{ \label{sc}\includegraphics[width=.27\textwidth]{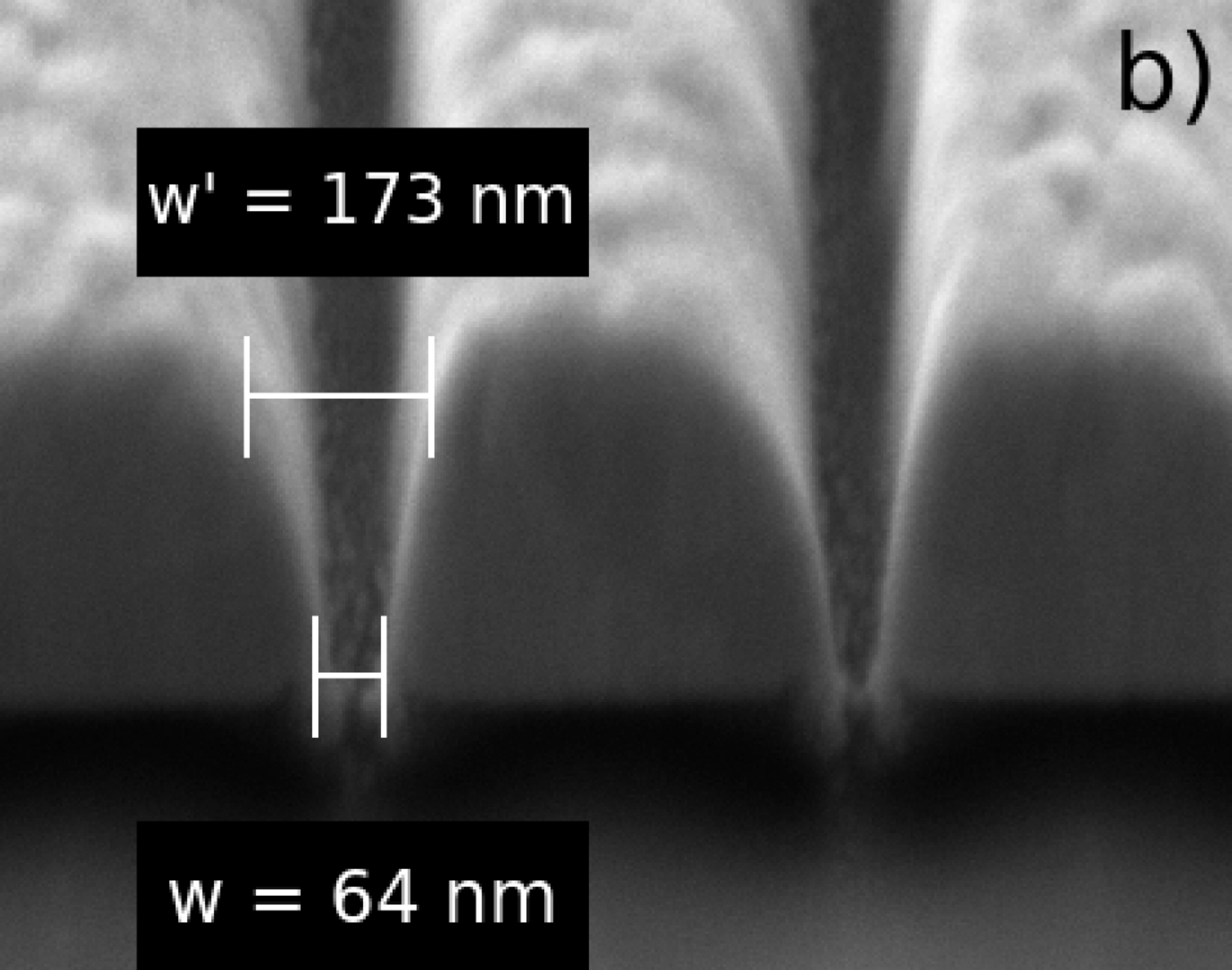}} 
\hfill
\subfloat[]{\label{pol_geometry} \includegraphics[width=.43\textwidth]{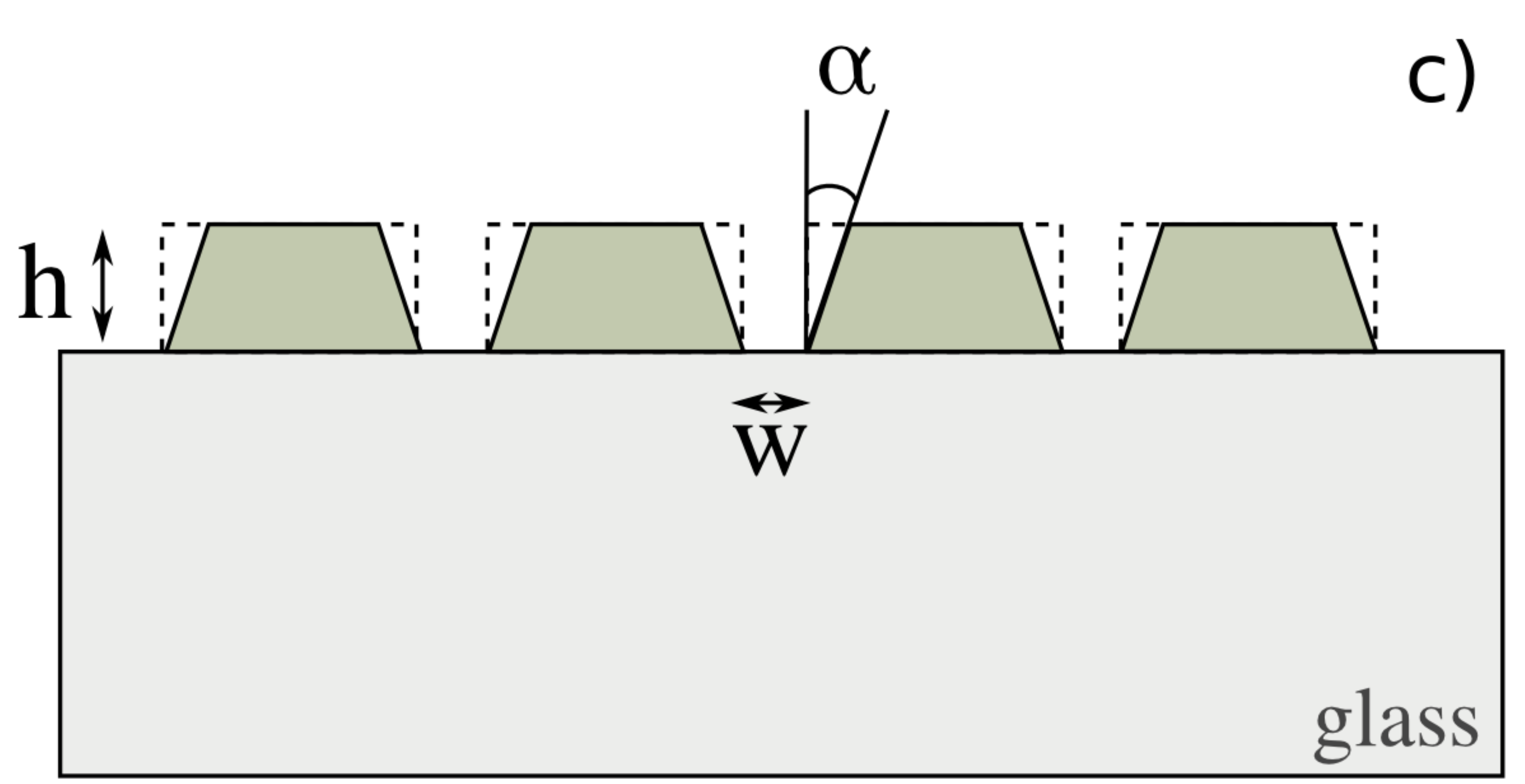}} 
\vspace{-15pt}
\caption{(a) Top and (b) side-view SEM images of the gold stripes. (c) Geometrical model taking into account the  trapezoidal shape of the wires.}
\label{sem1}
\end{figure}

 According to the simulations, an extinction ratio above 2,000 should be obtained for slit widths below 120$\,$nm. The first sample was fabricated with this largest possible width (Samples $A$, see Table \ref{oldsamples}), yet the measured values did not exceed 1:650. In order to achieve the desired performances, new samples with smaller slit widths were manufactured. Although the experimental ER slightly increases for $w=80 \,$nm, the simulations indicate a much stronger increase by at least a factor of 10. Moreover, the measured ER shows a large scatter for different samples  although the gratings exhibit a similar geometry and roughness in top view SEM pictures, typically similar to Fig. \subref*{st}. A significant difference appears when investigating the cross-section of the gold stripes using FIB milling. The side-view SEM pictures in Fig. \subref*{sc} now reveal a rather trapezoidal profile, with a non negligible difference between the lower and upper base lengths of about 100$\,$nm. This shape results from side redeposition of the ablated material during the milling of the stripes, which can be potentially reduced by using a multi-pass FIB writing mode.

\section{Grating optimisation with realistic geometrical model}

In order to understand the often reported reduced performances of the manufactured WGP compared to predictions, a deeper study involving further simulations has been carried out. To account for the observed geometry we introduce a refined model based on a more realistic trapezoidal stripe shape. The parameter $w$ now denotes the slit width at the bottom of the stripes, close to the substrate, and the angle $\alpha$ corresponds to the deviation from a perfect rectangular structure, as illustrated in Fig. \subref*{pol_geometry}. 

\begin{figure}[h]
         \centering
\captionsetup[subfigure]{labelformat=empty}
\begin{minipage}{0.49\linewidth}
\centering
\subfloat[\label{er}]{ \includegraphics[width=\textwidth, height=4.5cm]{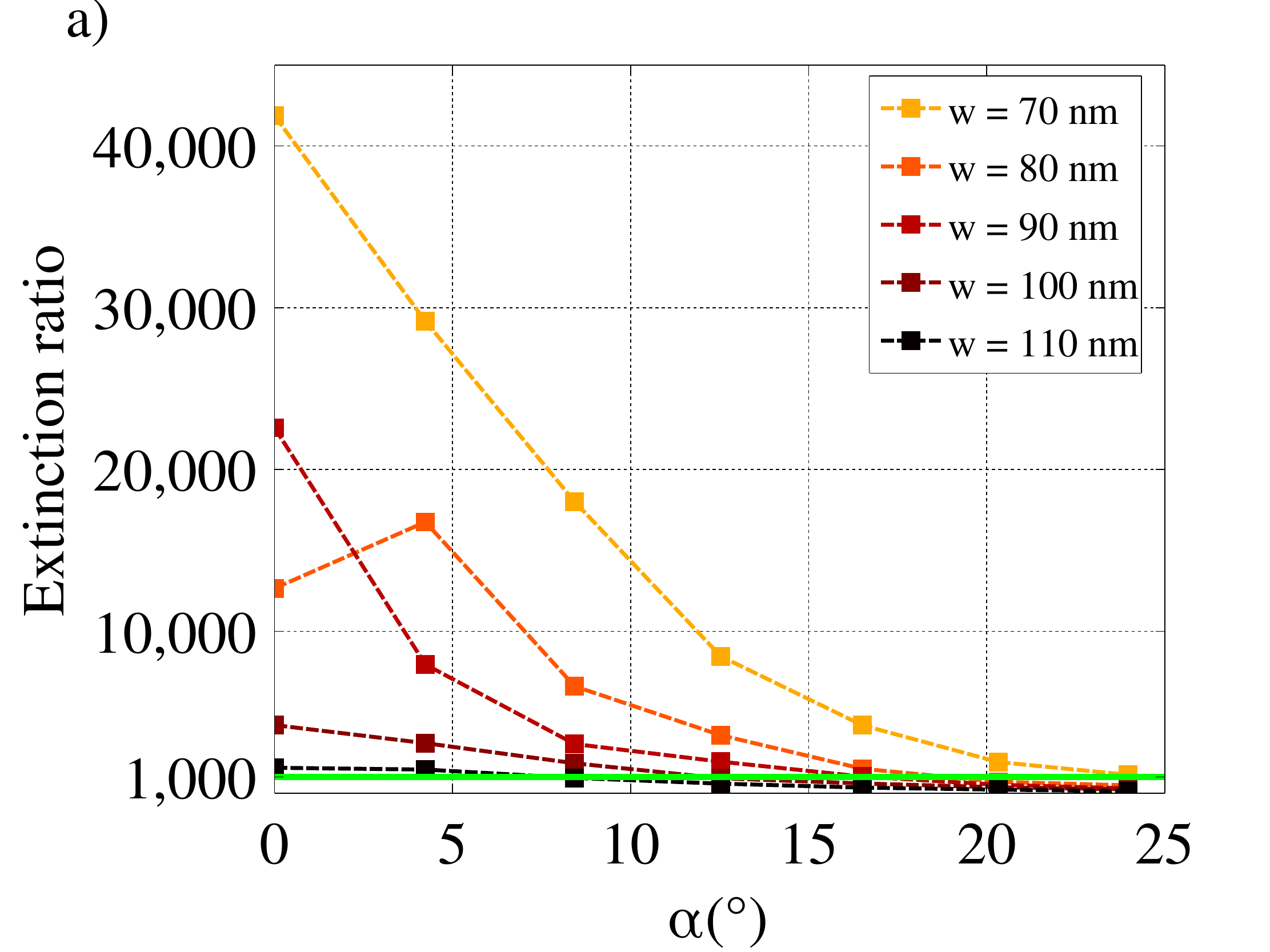}} 
\\ 

\hspace{.22cm} \subfloat[\label{tte}]{\includegraphics[width=.93\textwidth]{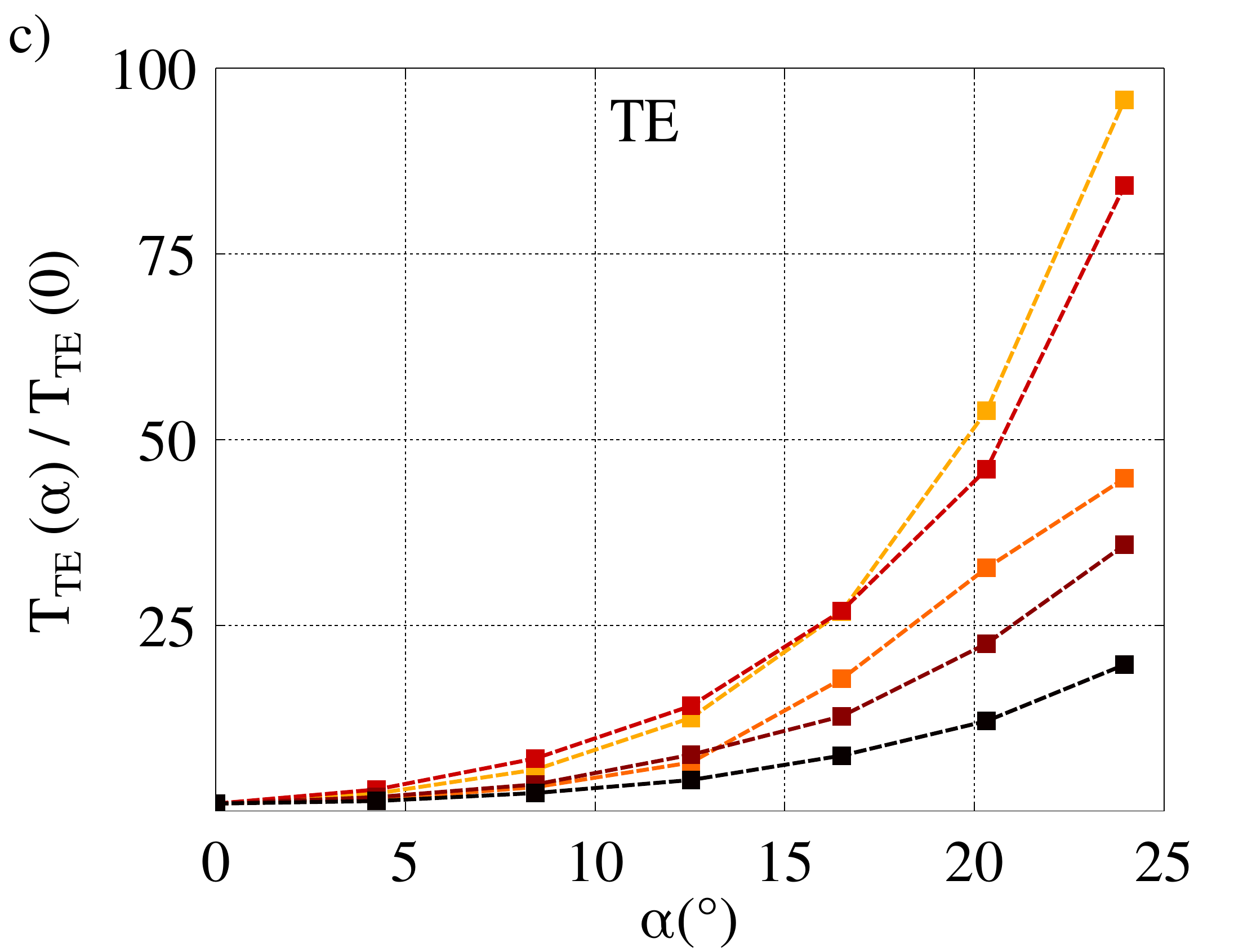}}
\end{minipage}
\hfill
\begin{minipage}{0.48\linewidth}
\subfloat[\label{ttm}]{\includegraphics[width=.92\textwidth, height=4.45cm]{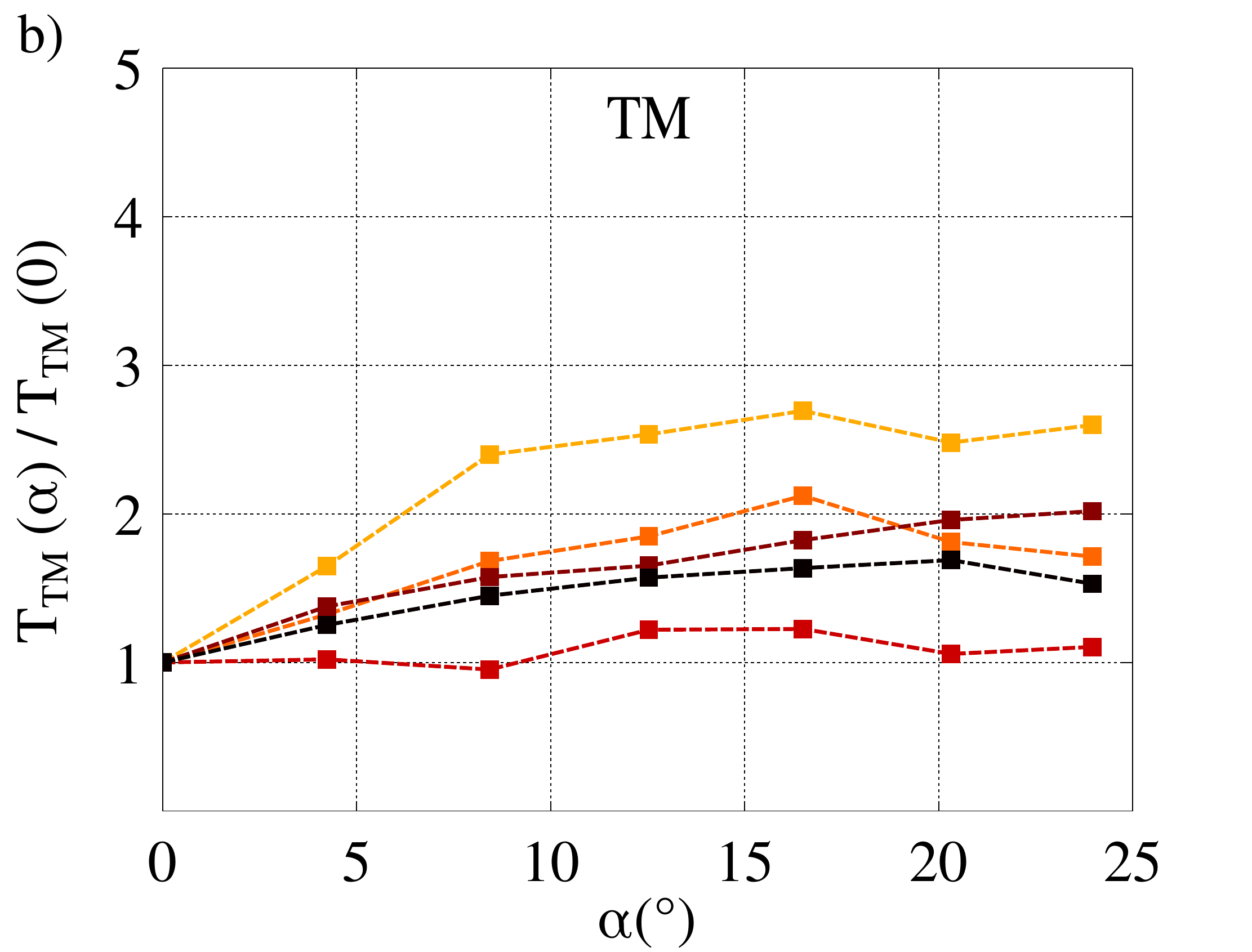}} \vspace{-6pt}

\subfloat[\label{weff}]{\includegraphics[width=.935\textwidth, height=4.6cm]{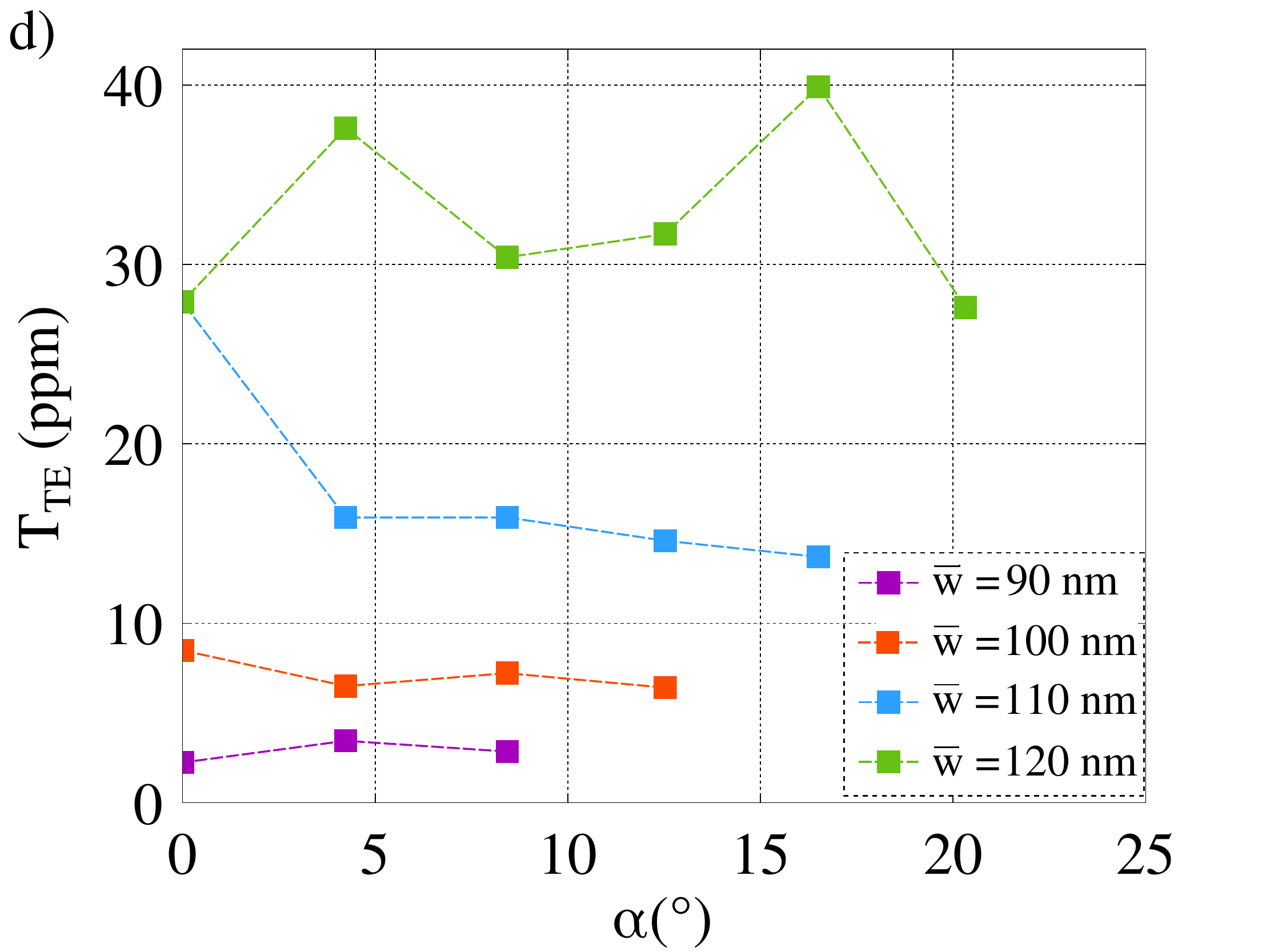}}
\end{minipage}
\vspace{-15pt}
\caption{Dependence of the performance on the slit angle $\alpha$ ($h = 270 \, $nm): (a) extinction ratio, (b),(c) transmission of \textit{TM} and \textit{TE} polarisation modes, respectively. The exponential increase of the transmission of \textit{TE}-modes is mainly responsible for the significant reduction of the extinction ratio. (d) Transmission of the \textit{TE} polarisation as a function of the effective slit width $\bar{w}$.}
\label{h270_trap}
\end{figure}

The transmission of both \textit{TE} and  \textit{TM} modes, as well as the ER are computed varying again the slit width $w$ and additionally the opening angle $\alpha$. Figure \subref*{er} shows an exponential decrease of ER when varying the angle $\alpha$, but also that the trapezoidal shape leads to slightly higher transmissions for the \textit{TM} modes, see Fig. \subref*{tte}. This is consistent with the case for rectangular shape, where $T_{TM}$ increases with $w$. This effect is nonetheless reduced in the case of the resonance ($w= 90 \,$nm). While the influence of $\alpha$ on \textit{TM}-states is limited, it has dramatic consequences for the \textit{TE}-polarisation, as can be seen in Fig. \subref*{ttm}. In this case we observe an exponential increase of the transmission with the opening angle, which consequently leads to the exponential decrease of the ER. When using the trapezoidal shape in the simulations, the theoretical ER values are reduced from 12700 (rectangular stripes, $w=120\,$nm) down to 500 for $B_1$ ($\alpha_{exp} = 25 \, ^o$) and 2700 for $B_2$ ($\alpha_{exp} = 16 \, ^o$). This trapezoidal shape indeed explains the order of magnitude discrepancy observed between simulations of perfectly rectangular stripes and realistic samples.

\begin{figure}[h]
         \centering
\captionsetup[subfigure]{labelformat=empty}
\centering
\begin{minipage}[c]{0.55\linewidth}
\subfloat[\label{fields}]{\includegraphics[width=\textwidth, height=4cm]{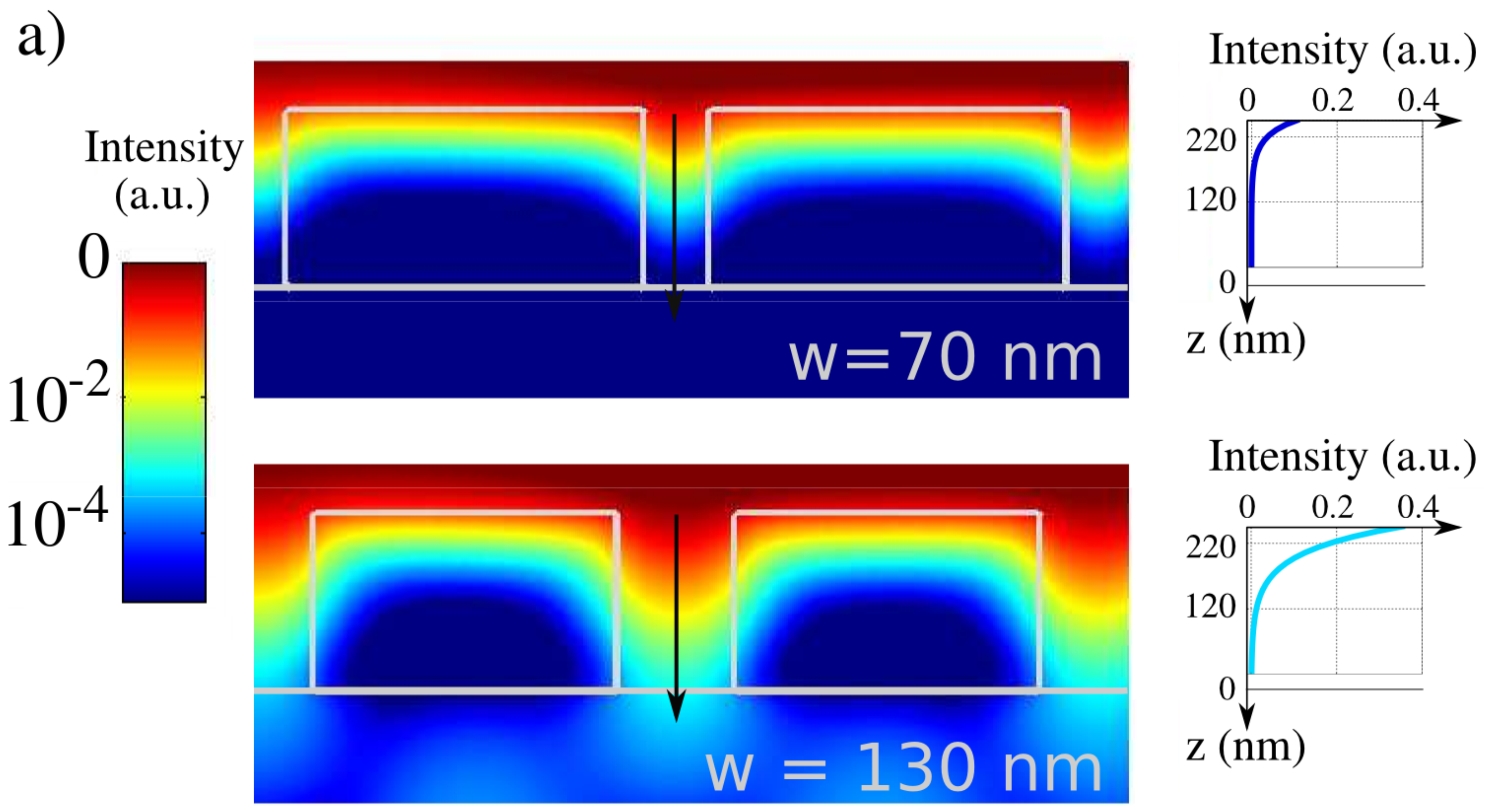}} 
\end{minipage} \hfill
\begin{minipage}[c]{0.42\linewidth}
\subfloat[\label{decay}]{\includegraphics[width=\textwidth]{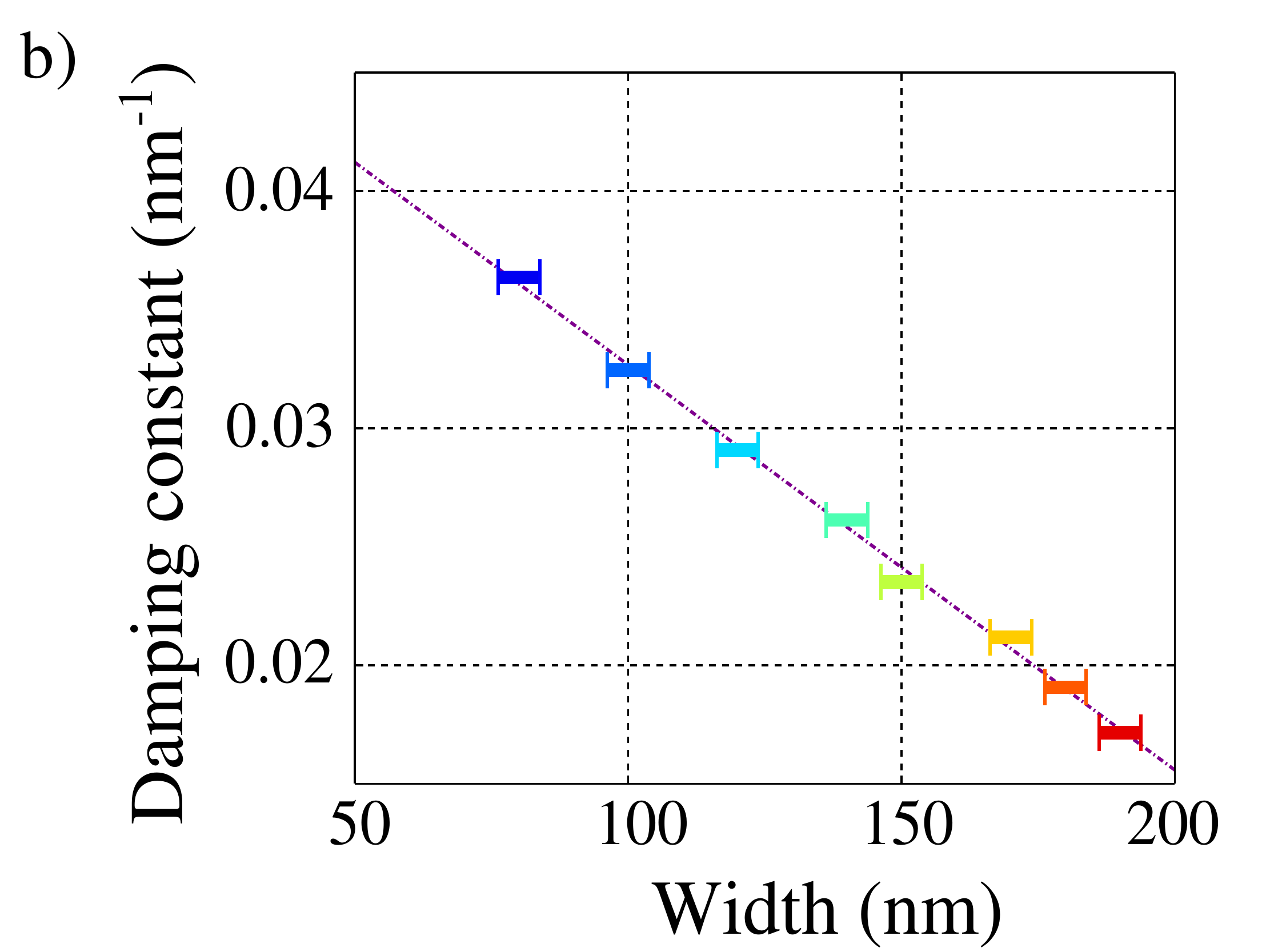}} 

\end{minipage}
\vspace{-15pt}
\caption{(a) Simulated intensity distribution for rectangular stripes separated by $w=70\, $nm and $w=130\,$nm respectively. The small insets show the exponential damping of the intensity within the slit ($I(z) \propto e^{- \gamma z}$). (b) Damping constant $\gamma$ extracted from the intensity profile for different slit widths.}

\label{field}
\end{figure}

 Since $T_{TM}$ is largely independent of the angle $\alpha$, it is the
dependence of the transmission of the TE-mode $T_{TE}$ which determines
the $ER$. A closer inspection adopting the model of guided waves in the slit  \cite{Snyder1983} shows that the \textit{TE}-mode is exponentially damped as the wavelength of the incoming field is above the cut-off wavelength. Figure \subref*{fields}  shows exemplarily the
result of the numerical simulations for two different heights.
Evidently, the field is exponentially damped when propagating through
the slit, which also explains the exponential dependence of the
transmission on the metal thickness. Analysing the dependence of the damping
coefficient on the slit width shows an approximately linear behaviour in
the region of relevant slit widths in Fig. \subref*{decay}, which is also in good agreement with
calculations of effective refraction indices in previous studies \cite{Schouten2003}. Given smooth variation of the slit width, $T_{TE}$ and thus ER can be well
approximated taking the effective, i.e., averaged, width of the slits into account. Figure \subref*{weff} shows that the unnormalised values of $T_{TE}$ for different average slit widths $\bar{w}= (w_{max}+w_{min})/2$ are indeed approximately independent of the angle $\alpha$.

\begin{figure}[!h]
         \captionsetup[subfigure]{labelformat=empty}
\centering
\begin{minipage}[c]{0.5\linewidth}
\subfloat[\label{array}]{\includegraphics[width=\textwidth]{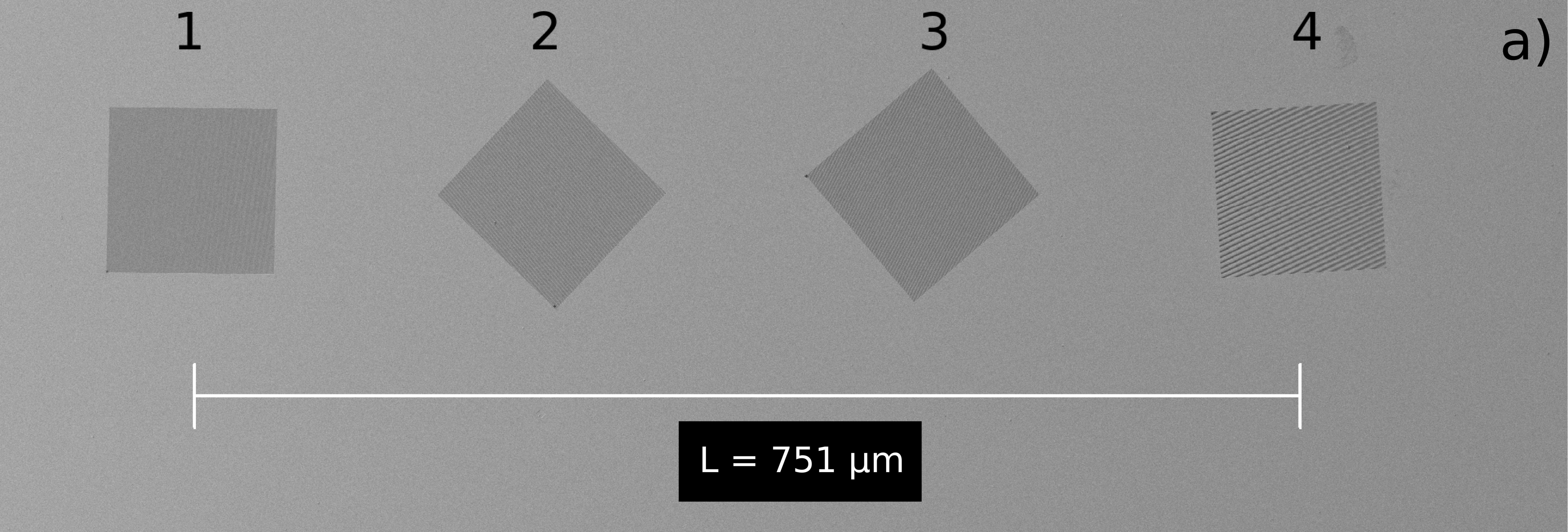}} 
\end{minipage}
\hspace{1cm}
\begin{minipage}[c]{0.289\linewidth}
\subfloat[\label{array2}]{\includegraphics[width=\textwidth]{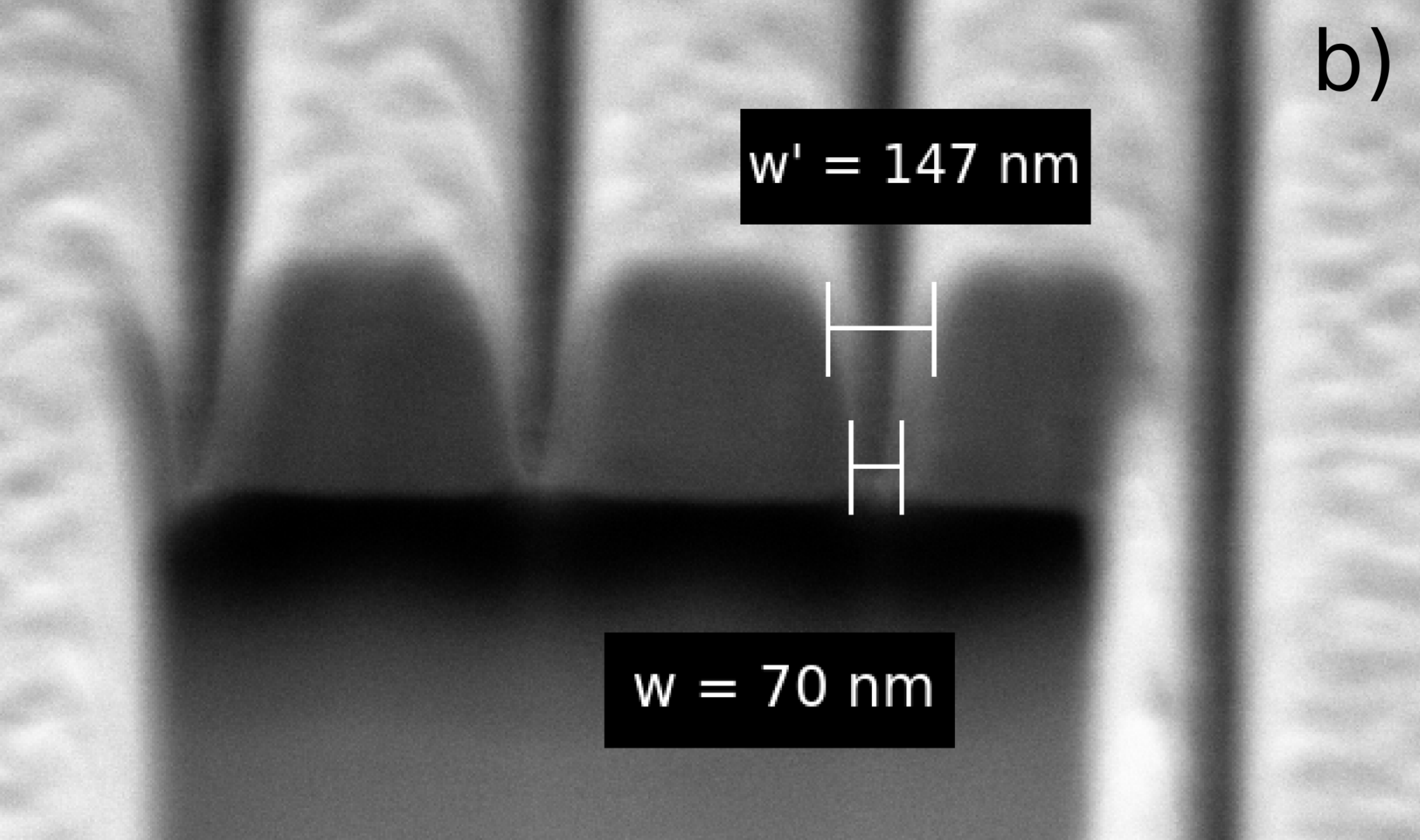}} 
\end{minipage}
\vspace{-15pt}
\caption{SEM pictures of a four-polariser array exhibiting extinction ratios up to 1,800. (a) Top view of the matrix. (b) Cross-section of the fourth grating. The decrease in performances associated with a large tilting angle $\alpha > 16 \, ^o$ was compensated by reducing the slit width.}
\label{sem2}
\end{figure}

Using these findings for the next optimisation step, we note that the targeted extinction ratio of 1,000 can be achieved for  rectangular stripes with $ w \leq 110 \,$nm, but not for achievable angles $\alpha \geq 16\, ^o$. Yet, reducing the slit width down to $70\,$nm (corresponding to $\bar{w} = 110\,$nm) brings a clear improvement. This simulation was verified experimentally by the fabrication of a new polariser array, presented in Fig. \ref{sem2}. Table \ref{ER_array} presents the characterisation of these four samples and compares them to the theoretical performances simulated with the refined model. As expected, the ER exceeds 1,000 and even reaches 1,800, yielding the best ER observed so far for 850$\,$nm. The transmission is similar for all samples and reaches 9$\,$\%. The simple simulation model based on trapezoidal stripes is therefore suited for realistic devices as it exhibits excellent agreement with the experimental data. 

 The remaining difference between experiment and theory is most likely due to the spatial resolution of the simulation and the measurement uncertainty of $\alpha$ on the SEM image, as well as the aforementionned effect of the adhesion layer. According to recent studies \cite{Ryu2008}, surface roughness should not degrade the performance significantly. In addition, the trapezoidal shape smoothens the field distribution compared to rectangular shapes, further reducing the influence of surface irregularities. \\

\begin{table}[sh]
\renewcommand{\arraystretch}{1.3}
 \centering
   \caption{Experimental results obtained after optimisation of the geometry using a trapezoidal model. The data exhibit clearly improved agreement with the simulations.}
    \vspace{-5pt}	
 \begin{tabular}{ c c c c c c}
Sample & $w$ (nm) &  $w_{max}$ (nm) &  $\alpha$ ($^o$) & ER (experimental) & ER (simulated) \\ \hline  \hline
1 & 70 & 150 & 16 & 1800 & 4200 \\ 
2 & 70 & 160 & 19 & 1620 & 2870 \\ 
3 & 80 & 160 & 16 & 1200 & 1510 \\ 
4 & 70 & 175 & 21 & 1150 & 1544 \\
 \end{tabular}
  \label{ER_array}
\end{table}

\section{Conclusion}

We presented clear evidence of the impact of geometrical deviations from perfect rectangular cross-section onto the performance of wire-grid-polarisers. Transversal SEM pictures indicate a trapezoidal shape of the stripes, a defect present with all currently used fabrication techniques. While better rectangular stripe profiles could be achieved, there will always be some imperfections left. Nevertheless, the real performance of wire-grid polarisers can be well simulated when accounting for the true shape of the wires or in a first approximation for the effective slit width. High extinction ratios can be achieved even with imperfect structures when optimising other parameters. Comparison between experiments and simulations shows, for the first time, significantly improved agreement, with experimental devices reaching extinction ratios of up to 1,800 and transmission of $9\,$\% at 850$\,$nm.

\section*{Acknowledgements}
 
The authors acknowledge technical help from P. Altpeter (LMU), P. Weiser and S. Matich (ZNN) regarding the fabrication of the polarisers . This project was funded by the excellence cluster Nano-Initiative Munich (NIM) and by the European projects CHIST-ERA/QUASAR, FP7/QWAD and FP7/CIPRIS.

\end{document}